\documentclass{article}
\usepackage{mlspconf}
\usepackage{amsmath}
\usepackage{mathtools}
\usepackage{amssymb}
\usepackage{graphicx}
\usepackage[colorlinks=false]{hyperref}
\usepackage{lipsum}
\usepackage{booktabs}
\usepackage{enumitem}
\usepackage[labelformat=simple]{subcaption}

\newcommand{\Lagr}{\mathcal{L}}
\DeclareMathOperator*{\softmax}{soft\!\max}
\DeclareMathOperator*{\argmin}{arg\!\min}

\def\thebibliography#1{%
  \section{References}  
  \list
    {[\arabic{enumi}]} 
    {\settowidth\labelwidth{[#1]}%
     \leftmargin\labelwidth%
     \advance\leftmargin\labelsep%
     \usecounter{enumi}%
     \setlength{\itemsep}{1pt} 
     \setlength{\parsep}{0pt}  
    }%
  \def\newblock{\hskip .11em plus .33em minus .07em}%
  \sloppy
  \clubpenalty4000
  \widowpenalty4000
  \sfcode`\.=1000\relax
  \ninept  
}

\copyrightnotice{979-8-3503-2411-2/25/\$31.00 {\copyright}2025 IEEE}

\toappear{2025 IEEE International Workshop on Machine Learning for Signal Processing, Aug.\ 31-- Sep.\ 3, 2025, Istanbul, Turkey}

\begin{document}

\title{Unseen Speaker and Language Adaptation for Lightweight Text-To-Speech with Adapters}


\name{%
   Alessio Falai$^{\star}$%
   \qquad Ziyao Zhang$^{\star}$%
   \qquad Akos Gangoly$^{\star}$%
}
\address{%
   $^{\star}$ Amazon AGI \\%
   \texttt{\{falai,zhaziyao,gangakos\}@amazon.com}
}

\maketitle

\begin{abstract}
In this paper we investigate cross-lingual Text-To-Speech (TTS) synthesis through the lens of adapters, in the context of lightweight TTS systems. In particular, we compare the tasks of unseen speaker and language adaptation with the goal of synthesising a target voice in a target language, in which the target voice has no recordings therein. Results from objective evaluations demonstrate the effectiveness of adapters in learning language-specific and speaker-specific information, allowing pre-trained models to learn unseen speaker identities or languages, while avoiding catastrophic forgetting of the original model's speaker or language information. Additionally, to measure how native the generated voices are in terms of accent, we propose and validate an objective metric inspired by mispronunciation detection techniques in second-language (L2) learners. The paper also provides insights into the impact of adapter placement, configuration and the number of speakers used.
\end{abstract}

\vspace{-5pt}
\section{Introduction}
\label{sec:introduction}
The landscape of Artificial Intelligence (AI) has recently witnessed significant strides in Text-To-Speech (TTS) synthesis \cite{tortoise}. Contemporary TTS model architectures predominantly rely on scaling laws to improve speech synthesis quality, albeit at the expense of larger model sizes \cite{valle}. Conversely, there remains a relatively sparse exploration of lightweight TTS approaches \cite{lightspeech,mbmelgan}. Going forward, due to limitations in terms of internet access or strict privacy requirements,  we believe that the demand for on-device speech synthesis will increase, which would require real-time and on-premise TTS models.

Moreover, as the demand for more personalised and versatile conversational agents grows, the need for adaptable models capable of handling diverse linguistic contexts becomes paramount. This ties in with the heated research avenue of cross-lingual voice cloning \cite{vallex} in TTS systems, where the goal is to transfer voices across languages. For example, some scenarios would require synthesising Mandarin speech using an English speaker's voice, where the English speaker does not have any Mandarin speech data. Existing solutions tackle this problem by training a large multi-speaker multilingual model with non-polyglot speakers (i.e. each speaker only speaks one language) \cite{zhang22c_interspeech,zhang19e_interspeech,nachmani2019unsupervised}. Given a large-enough number of examples, such models are able to disentangle speaker and language representations, thus enabling inference for unseen combinations of speaker and language. Other approaches involve the use of a Voice Conversion (VC) model as either a pre-processing or post-processing step. In the former case, cross-lingual VC is used to generate synthetic data of a target speaker speaking in a target language. Then, such synthetic data are used to train a single-speaker monolingual TTS system \cite{tts-synthetic-data-google}. In the latter case, cross-lingual VC is used to convert the speaker identity of a speaker in the target language, modeled by a TTS system, into the target speaker identity \cite{ellinas2022crosslingual}.

Outside of TTS, the realm of Natural Language Processing (NLP) has also experienced notable progress. Within the vast landscape of models, adapters \cite{bottleneck-adapters, unified-peft} have emerged as a crucial technique for achieving efficient fine-tuning of pre-trained Large Language Models (LLMs) \cite{gpt3}. Serving as compact, learnable parameter modules plugged on top of existing neural network architectures, they enable targeted adaptation without necessitating extensive retraining. As the quest for efficient and robust NLP solutions continues to evolve, different adapter solutions have been proposed for task adaptation \cite{prefix-tuning}, language adaptation \cite{madx} and more \cite{lora}.

In this work, we propose to look at cross-lingual voice cloning through the lens of adapters. We believe that adapters, when introduced to TTS systems, present a promising path for addressing one of the key challenges of adaptable systems, namely catastrophic forgetting \cite{catastrophic-forgetting}. Moreover, the concept of adapters has been already introduced to the TTS community, especially for parameter-efficient few-shot speaker adaptation \cite{tts-adapters-google, tts-adapters-nvidia, tts-adaptermix}. In this paper, our focus lies in demonstrating the feasibility and efficacy of adapters in the context of lightweight TTS architectures, for both cross-lingual speaker adaptation and language adaptation. 

The main contributions of this paper are as follows.
\begin{enumerate}[itemsep=2pt, topsep=2pt, parsep=2pt]
    \item We are the first to explore adapters in the context of lightweight, end-to-end (E2E) TTS models in a Generative Adversarial Network (GAN) training setting.
    \item We are the first to propose the use of adapters for TTS language adaptation, where a single-speaker monolingual model is fine-tuned to learn a new language.
    \item We propose a new objective method to evaluate accent similarity of a TTS systems against recordings of a native speaker in a target language.
\end{enumerate}

The rest of the paper is organised as follows. In Section \ref{sec:methodology} we introduce the backbone TTS model and the type of adapters we employ. In Section \ref{sec:metrics} we present the objective metrics we rely on to evaluate our models. Section \ref{sec:experiments} contains details about our experiments and results, while Section \ref{sec:conclusions} includes discussion points and concluding remarks.

\vspace{-5pt}
\section{Model backbone and adapters}
\label{sec:methodology}

\begin{figure*}
     \centering
     \begin{subfigure}[b]{0.19\textwidth}
         \centering
         \includegraphics[width=\textwidth]{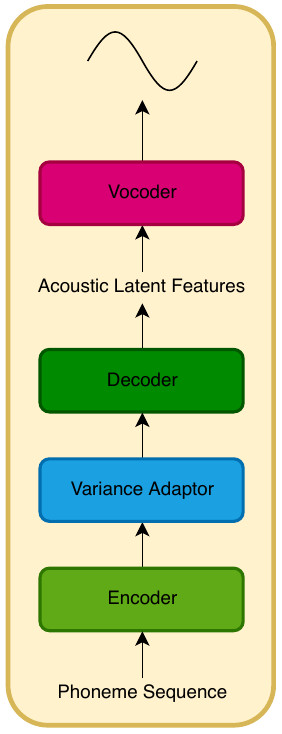}
         \caption{Entire model}
         \label{fig:arch-overall}
     \end{subfigure}
     \begin{subfigure}[b]{0.19\textwidth}
         \centering
         \textbf{\small{Variance Adaptor}}
         \includegraphics[width=\textwidth]{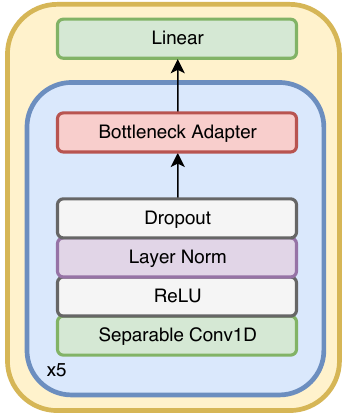}
         \textbf{\small{Encoder/Decoder}}
         \includegraphics[width=\textwidth]{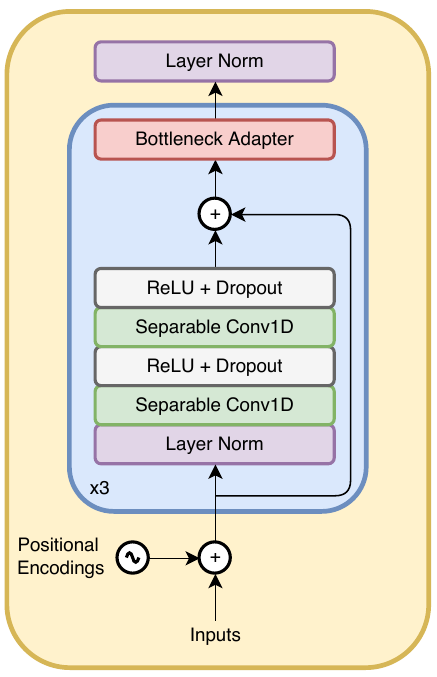}
         \caption{Acoustic model}
         \label{fig:arch-acoustic-model}
     \end{subfigure}
     \begin{subfigure}[b]{0.19\textwidth}
         \centering
         \includegraphics[width=\textwidth]{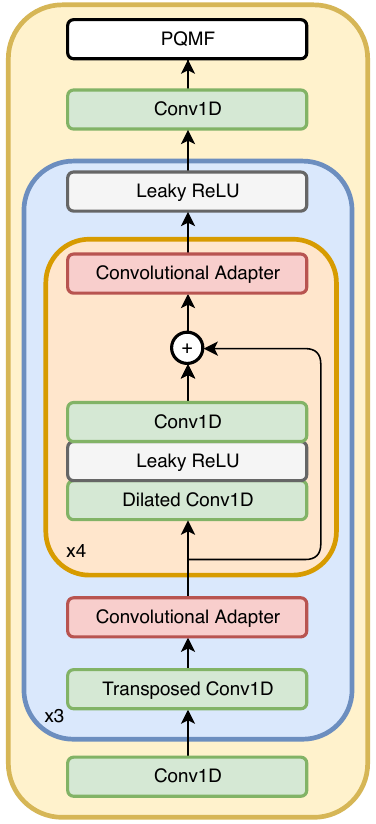}
         \caption{Vocoder}
         \label{fig:arch-vocoder}
     \end{subfigure}
     \begin{subfigure}[b]{0.19\textwidth}
         \centering
         \textbf{\small{Bottleneck Adapter}}
         \includegraphics[width=\textwidth]{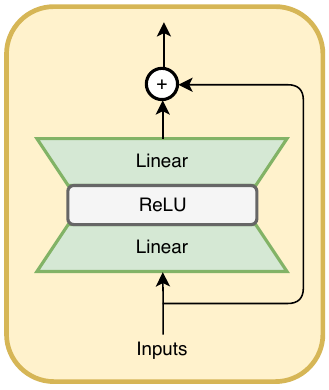}
         \textbf{\small{Convolutional Adapter}}
         \includegraphics[width=\textwidth]{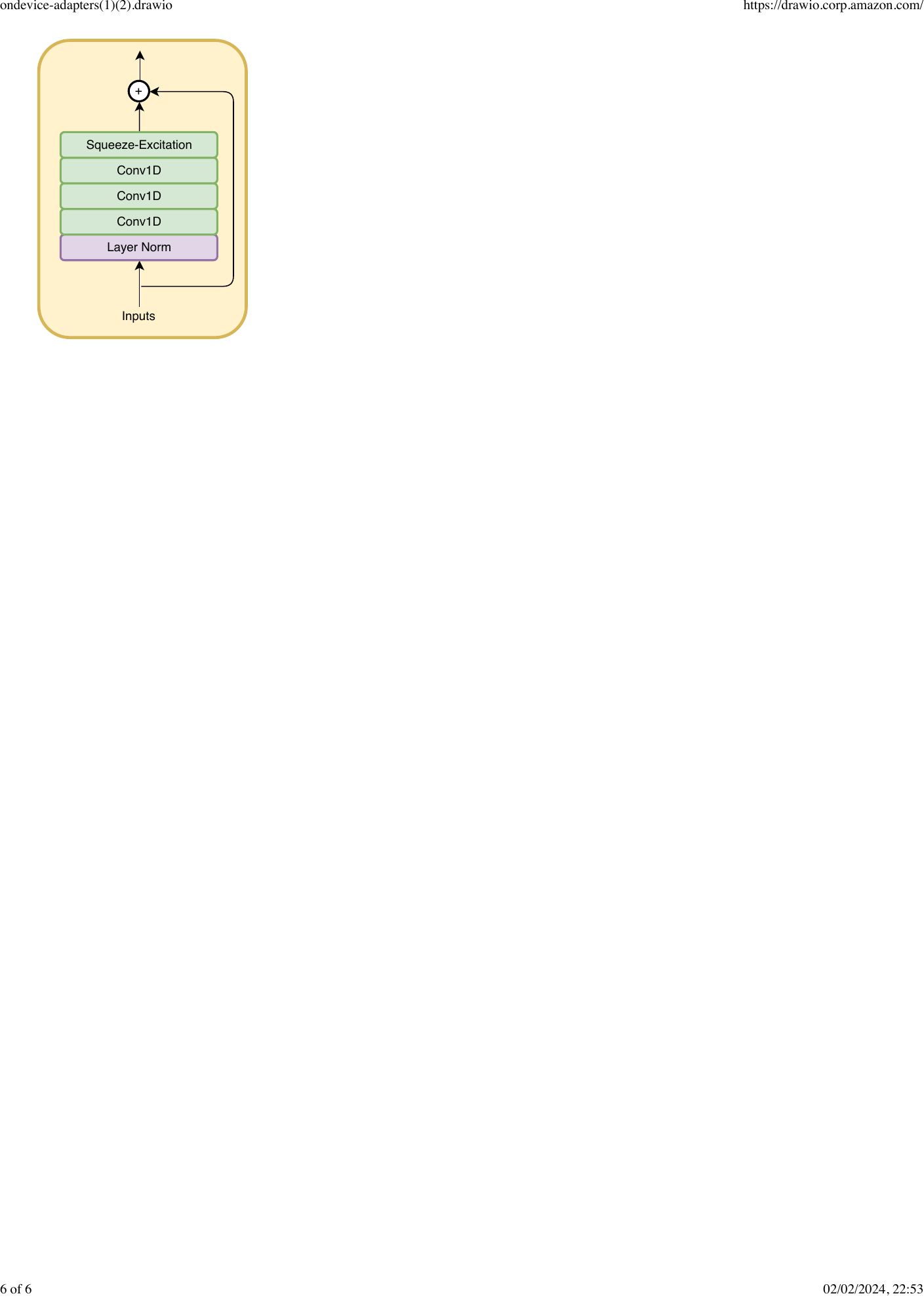}
         \caption{Adapters}
         \label{fig:arch-adapters}
     \end{subfigure}
    \caption{Model architecture at inference time. At training time, discriminators are also present.}
    \label{fig:architecture}
\end{figure*}

The base architecture used in this work is the one presented in \cite{vecino2023}, which is named Lightweight E2E-TTS (LE2E) and is shown in Fig.~\ref{fig:architecture}.

\subsection{Model description}
\label{subsec:model}
The model follows a GAN-based setup with a generator $G$ consisting of an acoustic model and a neural vocoder. The acoustic model, inspired by LightSpeech \cite{lightspeech}, generates unsupervised acoustic latents from phonemes and positional embeddings. It includes a text encoder, variance adaptor, and acoustic decoder. The neural vocoder, based on Multi-Band MelGAN (MB-MelGAN) \cite{mbmelgan}, converts these latents into waveform signals using up-sampling blocks and residual blocks. The discriminator set $D_k$ includes Multi-Period Discriminators (MPD) \cite{hifigan} and Multi-Resolution Discriminators (MRD) \cite{univnet}.

The model is optimised to minimise the total loss in \eqref{eq:total_loss} and the discriminator loss in \eqref{eq:discriminator_loss}, where the main adversarial training objective, expanded in \eqref{eq:discriminator_loss}, follows the least squares loss functions for non-vanishing gradient flows \cite{melgan}.
\begin{equation}
    \begin{split}
    \Lagr &= \Lagr_{dur} + \Lagr_{f0} + \Lagr_{G} + \lambda_{FM}\Lagr_{FM} + \\ 
    & + \lambda_{mel}\Lagr_{mel} + \lambda_{STFT}\Lagr_{STFT}
    \end{split}
    \label{eq:total_loss}
\end{equation}
\begin{equation}
    \begin{split}
    \Lagr_{G}(\hat{x}) &= \min_{G} \mathbb{E}_{\hat{x}} \left[ (D_{k}(\hat{x} - 1)^{2} \right] \\
    \Lagr_{D}(x, \hat{x}) &= \min_{D_k} \mathbb{E}_{x} \left[ (D_{k}(x - 1)^{2} \right] + \mathbb{E}_{\hat{x}} \left[ (D_{k}(\hat{x})^{2} \right]
    \end{split}
    \label{eq:discriminator_loss}
\end{equation}

In terms of variance modelling, $\Lagr_{dur}(d, \hat{d})$ refers to a frame-level Mean-Squared Error (MSE) between predicted $\Vec{d}$ and ground-truth durations $\hat{d}$ (extracted from a pre-trained forced aligner), while $\Lagr_{f0}(p, \hat{p})$ is a cross-entropy loss between predicted pitch bins $\softmax(p)$ and ground truth ones $\hat{p}$ (the result of 256-bin quantisation of standardised pitch).

Additional loss functions include a feature matching loss $\Lagr_{FM}$ \cite{melgan}, a multi-resolution STFT loss $\Lagr_{STFT}$ \cite{parallel-wavegan} and a mel-spectrogram loss $\Lagr_{mel}$ \cite{hifigan}.

\subsection{Adapters}
\label{subsec:adapters}
As introduced in Section \ref{sec:introduction}, adapters are a computationally-efficient solution to alter the performance of a pre-trained model towards a specific task or domain. Formally, consider a model with parameters $\theta$ (pre-trained and frozen) and $\phi$ (newly introduced). Efficient fine-tuning methods focus on optimising only $\phi$ based on loss $L$ and dataset $D$, as reported in \eqref{eq:adapters-optim}.

\begin{equation}
    \phi_{\star} \leftarrow \argmin_{\phi} L(D; \{\theta,\phi\})
    \label{eq:adapters-optim}
\end{equation}

Efficient fine-tuning techniques, such as bottleneck adapters \cite{bottleneck-adapters}, may insert parameters $\phi$ at various positions of the pre-trained model. Adapters include a down-projection matrix $W_{down}$, a non-linearity function $f$, an up-projection matrix $W_{up}$, and a residual connection $r$ to transform features $h$ into $h'$, which is formulated as $h' \leftarrow W_{up}\cdot f(W_{down}\cdot h) + r$.

In this work, we make use of two types of adapters: bottleneck \cite{bottleneck-adapters} and convolutional \cite{conv-adapters}. For the bottleneck adapter we utilise the vanilla version of it, as shown in Fig.~\ref{fig:arch-adapters}, setting $f=ReLU$ and the bottleneck dimension to $b_{dim}=16$. Early experiments revealed that looser bottlenecks $b_{dim}>16$ were detrimental for speaker/language adaptation.

Convolutional adapters were introduced in \cite{conv-adapters} for extremely efficient task adaptation, where authors show that they outperform bottleneck adapters and achieve comparable performance to LoRA \cite{lora} and prefix tuning, with only $0.94\%$ of their trainable parameters, which fits well with our requirements for lightweight TTS. We follow the original implementation, as shown in Fig.~\ref{fig:arch-adapters}, using three 1D convolutional layers with kernel size $k=[3,5,3]$ (the one with $k = 5$ is a depth-wise convolution \cite{depthwise-sep-convs}) along with layer normalisation and a Squeeze-and-Excitation (SE) module \cite{squeeze-excitation}. The output of the SE layer is then summed to the residual connection. 

\vspace{-5pt}
\section{Evaluation metrics}
\label{sec:metrics}
In the TTS field, model evaluations are tipically done via subjective tests, such as MUSHRA \cite{recommendation2001bs}. It is well known that human-based evaluations through crowd-sourcing platforms are a costly, lengthy, and unreliable process where factors like cheaters are guaranteed to bias results \cite{zequeira2019intra,buchholz2011crowdsourcing,burmania2015increasing,jimenez2018influence}. Thus, to assess the quality of generated voices, we discard subjective metrics and only employ objective ones. All evaluations are conducted with $1000$ unseen utterances.

\subsection{Speaker similarity}
We compute the Speaker Embedding Cosine Similarity (SECS) \cite{casanova21b_interspeech} by comparing reference (i.e. voice of the target speaker) and synthetic audios using a d-vector speaker verification model \cite{ge2e} to gauge speaker similarity. SECS involves the computation of cosine similarity (in the $[-1,1]$ range, with $1$ indicating ``same speaker") between the embeddings of two audios that are extracted from a pre-trained speaker encoder.

\subsection{Signal quality, intelligibility, naturalness}
We evaluate signal quality, intelligibility, and naturalness using TorchAudio-squim \cite{squim} metrics. In particular, we use a combination of Perceptual Evaluation of Speech Quality (PESQ) and Scale-Invariant Signal-to-Distortion Ratio (SI-SDR) for signal quality; Short-Time Objective Intelligibility (STOI) for intelligibility; and NORESQA-MOS (which we will refer to as MOS) for naturalness.

\subsection{Accent nativeness}
For assessing the nativeness of speakers in a target language, we propose a new objective metric for more reliable and reproducible accent comparisons. We base this metric on an Automatic Speech Recognition (ASR) model. In the field of Computer-Assisted Pronunciation Training (CAPT), various ASR-based systems have been proposed for detecting mispronunciation in speech uttered by second-language (L2) learners \cite{yan2020end,lo2020effective}. We refer to this task as Mispronunciation Detection and Diagnosis (MDD) \cite{peng2021study}. There are different ways to tackle this problem, but the most effective ones directly transcribe phoneme-level symbols \cite{lo2020effective} from speech signals, which are then used for measuring mispronunciations.

The MDD task targeted at L2 learners aligns perfectly with our goal of accent evaluation. Therefore, in this paper we use objective metrics produced by such MDD models for comparing the nativeness of accent in generated speech from different systems. In particular, the MDD model we use is very similar to the one introduced in \cite{peng2021study}, which is based off of wav2vec2 \cite{wav2vec2}. Our approach entails stacking a classification head on top of a large pre-trained multilingual wav2vec2 model\footnote{\href{https://huggingface.co/facebook/wav2vec2-large-xlsr-53}{https://huggingface.co/facebook/wav2vec2-large-xlsr-53}} and fine-tuning it with a Connectionist Temporal Classification (CTC) loss for the phoneme recognition task. As a proxy for accent similarity, we calculate the Phoneme Substitution Rate (PSR), i.e. the percentage of phonemes in an utterance which are substituted with a phoneme different from the ground truth. Numerous prior studies \cite{8462635, li18o_interspeech} have shown that phoneme substitution is a key pattern in accented speech. Therefore, we consider it as a reasonable objective metric for measuring non-native accent, as a higher phoneme substitution rate correlates with higher level of non-native accent. 

We validate this approach by comparing our historical human accent MUSHRA test results with the results produced by our MDD model, which confirmed that we are able to produce consistent rankings amongst different systems as human testers did. To do so, we compare pairs of systems from completed MUSHRA tests and compute the PSR of both systems in the pair. If the PSR for the system that has the highest MUSHRA score in the pair is lower than that of the other system in the pair, we mark this as a successful classification from the MDD model. We compare more than $50$ pairs of systems in different languages (\texttt{de-DE}, \texttt{en-AU}, \texttt{en-GB}, \texttt{es-US} and \texttt{fr-CA}) and obtain an average accuracy of more than $90\%$.

\vspace{-5pt}
\section{Experiments and results}
\label{sec:experiments}
In this section, we formally introduce the tasks of speaker and language adaptation and present our experimental results on both when comparing full fine-tuning (i.e. without adapters, fine-tuning the entire model) and adapter-based fine-tuning scenarios (i.e. with adapters, only fine-tuning adapters). 

\begin{table*}[t]
    \small
    \centering
    \caption{Results for the language adaptation task.}
    \begin{tabular}{c|c|c|c|c|c|c}
        \toprule
        \toprule
                & \textbf{SECS} ($\uparrow$) & \textbf{PESQ} ($\uparrow$) & \textbf{SI-SDR} ($\uparrow$) & \textbf{STOI} ($\uparrow$) & \textbf{MOS} ($\uparrow$) & \textbf{PSR} ($\downarrow$) \\ \midrule
        $\star_1$ Full fine-tuning & $0.43\pm 0.04$ & $3.15\pm 0.26$ & $20.81\pm 2.65$ & $0.99\pm 0.00$ & $3.99 \pm 0.17$ & $\boldsymbol{0.56\pm 1.33}$   \\
        $\star_2$ Adapters fine-tuning & $0.51\pm 0.04$ & $3.29\pm 0.25$ & $19.72\pm 2.20$ & $0.99\pm 0.00$ & $\boldsymbol{4.36 \pm 0.14}$ & $1.42\pm 1.95$ \\
        $\star_3$ Less adapters in vocoder & $0.49\pm 0.05$ & $\boldsymbol{3.57\pm 0.26}$ & $\boldsymbol{21.75\pm 2.97}$ & $0.99\pm 0.00$ & $3.61\pm 0.42$ & $1.47\pm 2.08$ \\
        $\star_4$ Single \texttt{es-ES} speaker & $\boldsymbol{0.63\pm 0.04}$ & $3.38\pm 0.26$ & $21.34\pm 2.61$ & $0.99\pm 0.00$ & $4.23\pm 0.23$ & $5.81\pm 4.53$ \\
        \bottomrule
        \bottomrule
    \end{tabular}
    \label{table:results-lang-adaptation}
\end{table*}
\begin{table*}[t]
    \small
    \centering
    \caption{Results for the speaker adaptation task.}
    \begin{tabular}{c|c|c|c|c|c|c}
        \toprule
        \toprule
                & \textbf{SECS} ($\uparrow$) & \textbf{PESQ} ($\uparrow$) & \textbf{SI-SDR} ($\uparrow$) & \textbf{STOI} ($\uparrow$) & \textbf{MOS} ($\uparrow$) & \textbf{PSR} ($\downarrow$) \\ \midrule
        $\dagger_1$ Full fine-tuning & $0.70\pm 0.05$ & $2.51\pm 0.36$ & $15.96\pm 3.32$ & $0.98\pm 0.00$ & $3.87 \pm 0.37$ & $8.44\pm 5.41$ \\
        $\dagger_2$ Adapters fine-tuning & $0.62\pm 0.05$ & $\boldsymbol{3.28\pm 0.26}$ & $20.15\pm 2.12$ & $0.98\pm 0.00$ & $4.19 \pm 0.25$ & $\boldsymbol{0.75\pm 1.56}$ \\
        $\dagger_3$ Adapters in entire model & $\boldsymbol{0.78\pm 0.04}$ & $2.98\pm 0.35$ & $\boldsymbol{20.40\pm 1.82}$ & $\boldsymbol{0.99\pm 0.00}$ & $\boldsymbol{4.23\pm 0.22}$ & $6.27\pm 3.80$ \\
        $\dagger_4$ Single \texttt{es-ES} speaker & $0.58\pm 0.04$ & $1.16\pm 0.03$ & $0.30\pm 1.50$ & $0.79\pm 0.03$ & $2.53\pm 0.12$ & $43.68\pm 6.58$ \\
        \bottomrule
        \bottomrule
    \end{tabular}
    \label{table:results-spk-adaptation}
\end{table*}

\subsection{Adapters settings}
\label{sec:adapters-settings}
We rely on bottleneck adapters for the LightSpeech-based encoder, decoder, duration and pitch predictors. We place one adapter block after each convolutional layer, as shown in Fig.~\ref{fig:arch-adapters}, and we set the bottleneck dimension to $b_{dim}=16$ for all adapter blocks, which gives us $150K$ additional trainable parameters in the acoustic model. On the vocoder side, we opt for convolutional adapters and we position one block after each upsampling layer and residual block, for a total of $15$ blocks: $5$ for each upsampling block, $1$ after the transposed convolution and one after each of the $4$ residual blocks. This gives us another $50K$ additional trainable parameters, for a total of $200K$ parameters, which corresponds to $10\%$ of the original model's capacity ($2M$ parameters). We experimented with using bottleneck adapters in the entire model, but opted for convolutional ones for the vocoder to keep the number of parameters under control, given the number of adapter modules needed in the vocoder to achieve high-quality adaptation. 

\subsection{Training settings}
\label{sec:training-settings}
All backbone models (described in Sections \ref{subsec:language-adaptation} and \ref{subsec:speaker-adaptation}) underwent training for $1M$ steps, with an effective batch size of $128$ samples. The training utilized the AdamW optimizer \cite{adamw} with $\beta_1 = 0.8$ and $\beta_2 = 0.99$, along with an initial learning rate set at $2 \times 10^{-4}$ and an exponential decay scheduler with a factor of $\gamma = 0.99$. Additionally, a weight decay penalty factor of $1 \times 10^{-2}$ was applied. For a more in-depth understanding, readers are directed to \cite{vecino2023}.

During fine-tuning, models are optimised by freezing all modules with the exception of adapter blocks, the phoneme embedding matrix and the speaker/language encoding look-up-tables. Losses are the same as those used for training the backbone models and all fine-tuning experiments are run for $200$ epochs with the same optimiser and LR scheduler used for the backbone models, with the only difference in the initial LR for the full fine-tuning version and the adapter-based fine-tuning, equal to $1 \times 10^{-5}$ and $1 \times 10^{-4}$, respectively. This is because adapters contain newly introduced parameters and we find it beneficial to favour a higher degree of exploration. 

\subsection{Language adaptation}
\label{subsec:language-adaptation}
Given a TTS model pre-trained on speaker $S_1$ (where $S_1$ speaks in language $L_1$), we define language adaptation as adapting it to speak in a new, unseen language $L_{2}$ while maintaining the speaker identity of $S_{1}$. In this work, we pre-train the backbone model described in Section \ref{subsec:model} on roughly $10$ hours of recordings from a native British English (\texttt{en-GB}) speaker ($S_1$). Our goal is to adapt this pre-trained model to Castillan Spanish (\texttt{es-ES}) and we do so by fine-tuning the model on $100$ hours of \texttt{es-ES} data that spans across $n=25$ different speakers $S_2, \dots S_{n+1}$. The setup of adapters and the training settings are reported in Sections \ref{sec:adapters-settings} and \ref{sec:training-settings}.

In Table \ref{table:results-lang-adaptation} we show results comparing full fine-tuning ($\star_1$) and adapter-based fine-tuning ($\star_2$). What we observe is that adapter-based fine-tuning results in overall better signal quality, intelligibility and naturalness, as indicated by PESQ, SI-SDR, STOI and MOS metrics, higher speaker similarity to the target speaker $S_1$ and a slightly lower accent nativeness w.r.t. the full fine-tuning scenario. Informal listening shows that the lower accent nativeness in the adapter-based scenario is due to full fine-tuning overfitting on one of the \texttt{es-ES} speakers in the dataset, which explains the lower SECS score. 

To understand the effect of adapter placement on its efficacy, we reduce the number of convolutional adapters in the vocoder from $15$ to $3$, removing all adapters after residual blocks. In Table \ref{table:results-lang-adaptation} ($\star_3$) we can see a slight degradation for both speaker similarity and accent nativeness metrics. Similarly, in Table \ref{table:results-lang-adaptation} ($\star_4$) we show that relying on only $1$ \texttt{es-ES} speaker (instead of $n$) for fine-tuning (with adapters) the TTS model to language $L_2$ results in severe degradation in accent nativeness, with an expected improvement in speaker similarity (due to English accent leakage from speaker $S_1$).

\subsection{Speaker adaptation}
\label{subsec:speaker-adaptation}
Given a TTS model pre-trained on a set of $n$ speakers $S_2, \dots S_{n+1}$ (where $S_2, \dots S_{n+1}$ all speak in language $L_2$), we define speaker adaptation as adapting it to sound like a new, unseen speaker $S_1$ (which speaks in language $L_1$). We use the same dataset from Section \ref{subsec:language-adaptation}, but in the opposite way: the \texttt{es-ES} data is used for training the backbone model, while the \texttt{en-GB} data is used for adapting the speaker identity.

In Table~\ref{table:results-spk-adaptation} we show results comparing full fine-tuning and adapter-based fine-tunings for the speaker adaptation scenario. In particular, we train two model variants with adapters added to different modules, i.e. one with adapters added only in the vocoder ($\dagger_2$) and another one with adapters additionally added in the acoustic model ($\dagger_3$). Similar to before, we have as one comparison baseline the full-fine-tuning scenario without adapters ($\dagger_1$). Moreover, we pre-train the TTS model on only one es-ES speaker (instead of $n$) and then fine-tune it (with adapters) to speaker $S_1$ for exploring the effect of reducing the amount of training data in $L_{2}$ ($\dagger_4$).

Overall, we observe that adapter-based fine-tuning achieves the best results across all axes. Looking closer, we can see that removing adapters from the acoustic model leads to significantly better accent nativeness, reflected in almost ten-fold reduction in PSR. This indicates that the model encodes the majority of language-agnostic speaker information in the vocoder modules. All metrics considered, we conclude that vocoder-only adapter fine-tuning achieves the best outcome. On the other hand, it is also clear that reducing the amount of training data in $L_{2}$ during pre-training of the TTS model does not favour speaker adaptation. This suggests that the model must learn meaningful speaker representations in the pre-training phase for achieving high-quality speaker adaptation.

\vspace{-5pt}
\section{Conclusions}
\label{sec:conclusions}
In this paper, we explored the application of adapters in the context of E2E lightweight TTS, with a particular focus on cross-lingual speaker adaptation and language adaptation. Our investigation shows that adapters are a viable solution to adapt pre-trained TTS models to unseen languages or speakers, with only 10\% additional model parameters. Comparing speaker and language adaptations, we find the former to be an easier task for target models, even though the latter results in the best combination of objective metrics when the goal is to transfer voices across languages. Finally, our experiments reveal that both strategic positioning of adapters in the vocoder and speaker variety in the target language play a crucial role for accent nativeness and speaker similarity.

\vfill\eject

\end{document}